\documentclass{aa}
\usepackage{txfonts, graphicx, psfig, epsf, epsfig}

\newbox\grsign \setbox\grsign=\hbox{$>$}
\newdimen\grdimen \grdimen=\ht\grsign
\newbox\laxbox \newbox\gaxbox
\setbox\gaxbox=\hbox{\raise.5ex\hbox{$>$}\llap
        {\lower.5ex\hbox{$\sim$}}}\ht1=\grdimen\dp1=0pt
\setbox\laxbox=\hbox{\raise.5ex\hbox{$<$}\llap
        {\lower.5ex\hbox{$\sim$}}}\ht2=\grdimen\dp2=0pt
\newcommand{\simlt}{\mathrel{\copy\laxbox}}
\newcommand{\simgt}{\mathrel{\copy\gaxbox}} 
\newcommand{\xmm}{{\it XMM-Newton}}
\newcommand{\lum}{\thinspace\hbox{$\hbox{erg}\ \hbox{s}^{-1}$}}

\begin{document}

   \title{A second glance at SN~2002ap and the M\,74 field with \xmm}


   \author{Roberto Soria \inst{1},
	  Elena Pian \inst{2}
	  \and
	  Paolo A. Mazzali \inst{2}
          }

   \offprints{{\tt Roberto.Soria@mssl.ucl.ac.uk}}

   \institute{Mullard Space Science Laboratory, 
          University College London, Holmbury St Mary, 
          Surrey RH5 6NT, UK  \\
              email: {\tt{Roberto.Soria@mssl.ucl.ac.uk}}
	\and
	INAF, Osservatorio Astronomico di Trieste,       
	via Tiepolo 11, I-34131 Trieste, Italy \\
              email: {\tt{pian@ts.astro.it, mazzali@ts.astro.it}}
} 


   \date{Received  29 April 2003; Revised 04 September 2003  }

   \abstract{We have re-observed the field of M\,74 in January 2003 with \xmm, 
11 months after the X-ray detection of SN~2002ap. From a comparison 
of the two \xmm\ observations we obtain more accurate values for the X-ray luminosity 
and colours of the source five days after the event, and a limit on its decline rate. 
We compare its X-ray behaviour (prompt soft X-ray emission, relatively 
low luminosity) with that of other Type Ic SNe, and speculate that SN~2002ap 
may share some physical properties (low mass-loss rate and high-velocity stellar wind 
from the progenitor star) with the candidate hypernova/gamma-ray-burst 
progenitor SN~1998bw, 
but with a lower (non-relativistic) speed of the ejecta.
We suggest that the X-ray emission observed in 2002 is likely to come 
from the radiatively-cooling reverse shock, at a temperature $kT \approx 0.8$ keV, 
and that this soft component was already detected 5 d after the event 
because the absorbing column density of the cool shell between 
the forward and reverse shocks was only $\sim 10^{20}$ cm$^{-2}$, 
ie, the shell was optically thin in the soft X-ray band.
The new \xmm\ data also allowed us to continue monitoring two bright 
variable sources in M\,74 that had reached peak 
luminosities $> 10^{39}$ erg s$^{-1}$ 
in previous \xmm\ and {\it Chandra} observations. 
Finally, we used two {\it Chandra} observations 
from 2001 to investigate the luminosity and colour distribution of the X-ray source 
population of M\,74, 
typical of moderately-active late-type spirals. 
}

\authorrunning{R. Soria, E. Pian, \& P. A. Mazzali}
   \maketitle
%

\section{Introduction}

To date, only about 20 supernovae (SNe) have been detected in the
X-ray band (see Schlegel 1995; Immler et al.~1998; 
Schlegel 1999; Pian et al. 2000; Immler et al.~2001; 
Pooley et al. 2002; Schlegel 2001; Immler et al.~2002; 
and the recent review by Immler \& Lewin 2002). 
Most studies have been conducted in the soft X-ray band 
by {\it ROSAT} and more recently by {\it Chandra} and \xmm\ 
(Fox et al. 2000; Kulkarni \& Fox 2003;  Pooley \& Lewin 2002; Schlegel 2002; 
Zimmermann \& Aschenbach 2003).
Information at harder energies is much more limited: only SNe 1987A, 1993J and 1998bw have
been investigated at energies higher than 13 keV, and only the first two have been detected 
(Sunyaev et al.~1987; Inoue et al.~1991; Leising et al.~1994).

SN~2002ap in M\,74 (NGC 628) is one of only four Type~Ic SNe detected in the X-rays, the
others being the low-mass, normally energetic SN~1994I (Immler et al. 1998; Immler et al.
2002), the massive and energetic "hypernova" SN~1998bw (Galama et al. 1998; Iwamoto et al.
1998; Pian et al. 2000), and the normal SN~2003L (Boles et al.~2003; Kulkarni \& Fox 2003; 
Matheson et al. 2003).
From its broad optical spectral features and its high kinetic energy, Mazzali et
al.~(2002) argued that SN~2002ap can be classified as a hypernova, a class of SNe
characterised by an energetic, probably asymmetric explosion and by a large mass 
of the collapsing star (Paczy\'nski 1998; MacFadyen \& Woosley 1999). These
physical circumstances make hypernovae strong candidates to explain the origin of gamma-ray
bursts (GRBs). Indeed, SN~1998bw was observed in the {\it BeppoSAX} error box of GRB980425 and showed a
good temporal agreement with it. Hypernovae are also thought to be progenitors of
stellar-mass and possibly intermediate-mass black holes (BHs). Studying a Type Ic event at
all wavelengths, and especially at high energies, is therefore crucial to understand the
possible link between GRBs and SNe, and to test the identification of hypernovae as Type Ic
SNe.

The host galaxy of SN~2002ap is itself an interesting 
target for an X-ray study: it is a face-on (inclination angle 
$< 7^{\circ}$, Shostak \& van der Kruit 1984), late-type spiral 
(morphological type Sc), with star formation along 
well-defined arms (eg, Kennicutt \& Hodge 1980). Its distance 
remains uncertain: 
recent photometric measurements put it at $7.3$ Mpc 
(Sharina et al.~1996; Sohn \& Davidge 1996). 
Previous estimates, however, ranged from 2 to 20 Mpc 
(eg, Bottinelli et al.~1984; Sandage \& Tammann 1974). 
A distance of 8.8 Mpc was recently adopted  
by Huchra et al.~(1999) based on its redshift.
In this paper we shall assume a distance of 7.3 Mpc 
for all flux-to-luminosity conversions.

\section{Data analysis and Results}

\subsection{Log of the observations}

The field of SN~2002ap and its host galaxy M\,74 
were observed by all instruments on board \xmm\  
with two Target-of-Opportunity observations: 
the first on 2002 February 2.03--2.42 UT (revolution 394, 
less than 5 days after the SN event); the second 
on 2003 January 7.53--7.83 UT
(revolution 564). Both were taken 
in full frame, thin filter mode.
After rejecting intervals characterised by 
highly fluctuating background, we kept a good time interval 
of 21.5 ks for the 2002 pn observation, 
and 20.9 ks for the 2003 one. For the MOS, 
the good-time intervals were 23.7 ks in 2002 and 24.4 ks in 2003. 
We processed both datasets, and extracted spectra 
and lightcurves using version 5.4 of the XMM-Science Analysis 
Software ({\footnotesize{SAS}}); we considered only "pattern-0" events in both pn 
and MOS. We then used standard tools such as {\footnotesize{XSPEC}} 
(Arnaud 1996) for further data analysis.

In addition to the 2002 and 2003 \xmm\ observations, 
M\,74 was observed twice by {\it Chandra} ACIS-S: 
on 2001 June 19 (46.4 ks), and on 2001 October 19 
(46.2 ks). On both occasions, the back-illuminated S3 chip 
was used. We obtained the {\it Chandra} dataset 
from the public archive and analysed it with 
the standard {\footnotesize{CIAO}} software. 
We inspected the background count rates during 
the two exposures, and chose to retain both intervals 
in full (See also Krauss et al.~(2003)).
We used standard source-finding routines ({\it wavdetect} 
and {\it celldetect}, which give similar results) 
to identify the point sources. To increase the signal-to-noise 
ratio of faint sources, we also built a merged event 
file from the two {\it Chandra} observations, 
and used it to compile a source list and their average 
count rates in the full energy band ($0.3$--$8$ keV) 
and in three narrower bands (Table A.1).

\subsection{Count rates for SN~2002ap}

SN~2002ap is not detected in the 2003 January observation:  
neither in the single pn and MOS images, nor 
in a combined EPIC image. By comparison with the detection limit of 
the faintest sources in the combined image, we estimate a 3 sigma 
upper limit to the EPIC pn count rate 
of $\approx 4 \times 10^{-4}$ cts s$^{-1}$
in the $0.3$--$12$ keV band.\footnote{All the count rates 
listed here and hereafter have already been corrected for 
the enclosed energy fraction and the vignetting.}

Instead, SN~2002ap is detected as an X-ray source in the 2002 February 
observation, both in the EPIC pn and in the EPIC MOS images. 
A preliminary estimate (Sutaria et al.~2002) yielded an observed flux of
$1.07^{+0.63}_{-0.31} \times 10^{-14}$ erg cm$^{-2}$ s$^{-1}$ in the 
$0.3$--$10.0$ keV band.
Sutaria et al.~(2002) obtained this value by extracting the events 
from a 40\arcsec\ circle centred on the optical position 
of SN~2002ap, then subtracting the contribution from 
the background and from an unrelated source (CXOU J013623.5+154458) 
located $\approx 15$\arcsec\ from the SN.
The contribution from CXOU J013623.5+154458 
was estimated from the archival {\it Chandra} ACIS observation 
of M\,74 taken on 2001 October 19.
However, this method may lead to an inaccurate estimate, 
given the large uncertainty in the spectrum 
of CXOU J013623.5+154458 (ACIS-S count rate of 
$5.7 \times 10^{-4}$ cts s$^{-1}$, ie, only 26 counts 
in the 46.2-ks {\it Chandra} observation).
Moreover, a 40\arcsec\ extraction radius 
includes a significant contribution from 
at least one other {\it Chandra} source (CXOU J013626.6+154458), 
located 38\arcsec\ from the SN (ACIS-S count rate of 
$1.9 \times 10^{-4}$ cts s$^{-1}$).

We obtained an improved estimate 
of the X-ray flux of SN~2002ap in two different ways. 
Firstly, we used a much smaller extraction region 
(12\arcsec\ radius), to reduce the contamination from 
CXOU J013623.5+154458 and CXOU J013626.6+154458. 
We used the latest {\footnotesize{SAS}} calibration files, 
to correct for the smaller energy fraction 
enclosed in this region ($\approx 60$\% of the total, 
for channel energies $\simlt 2$ keV). We used 
the {\footnotesize{SAS}} tasks {\tt rmfgen} and {\tt arfgen} 
to construct accurate response matrices at the position 
of the source. The background was extracted from the same 
pn chip, in a region that did not contain any detected sources.

Alternatively, we used the 2003 image as a background 
to be subtracted from the 2002 dataset. To make sure we could 
do that, we compared the background contribution  
in the good-time-intervals of the 2002 and 2003 EPIC 
datasets. The (low) background count rate in the 
$0.3$--$12$ keV range is consistent with being the same 
in both observations (differences are $< 2$\% for the pn, 
and $< 5$\% for the MOS). The pointing of the spacecraft 
was also approximately the same in 2002 and 2003, hence 
SN~2002ap is located in the same chip, nearly on-axis 
on both occasions. Comparing the \xmm\ and {\it Chandra} datasets, 
we could verify that the brighter (and closer) 
of the two {\it Chandra} sources, CXOU J013623.5+154458, 
does not show significant flux variations from 2001 to 2003.
Therefore, we assumed that its state in 2002 was similar to
that displayed in 2001 and 2003.
If we assume that other possible faint sources 
in the region around SN~2002ap did not vary between 
2002 and 2003, we can subtract the flux detected 
at the position of SN~2002ap in the 2003 observation 
from the value measured in 2002. This gives us at least 
a lower limit on the luminosity of SN~2002ap in 2002.

Applying the first method (ie, using the 2002 dataset only, 
for source and background), 
we obtain a pn count rate of $(2.75 \pm 0.57)\times 10^{-3}$ cts s$^{-1}$ 
in the $0.3$--$12$ keV band. By comparing the count rates 
in three separate narrow bands (0.3--1 keV, 1--2 keV, 2--12 keV) we notice that 
the source has a soft spectrum. 
In fact, the count rate in the $2$--$12$ keV band 
is affected by some residual 
contamination from the nearby hard spectrum source 
CXOU J013623.5+154458, and should be considered 
an upper limit to the hard X-ray emission from SN~2002ap.
For MOS1, the corrected count rate on 2002 February 2 
was $(7.0 \pm 2.7) \times 10^{-4}$ cts s$^{-1}$; 
for MOS2, $(7.8 \pm 2.9) \times 10^{-4}$ cts s$^{-1}$. 
The low number of source counts ($\approx 10$ counts in 
each of the two detectors) does not allow a significant 
colour determination from the MOS's.

We then applied the second method, analysing 
the difference in the emission at the position 
of SN~2002ap between 2002 and 2003. 
We used two different source extraction radii: 
for a 12\arcsec\ radius, we obtain that the pn count 
rate in 2002 was higher than in 2003 by 
$(2.6 \pm 0.5)\times 10^{-3}$ cts s$^{-1}$ 
in the $0.3$--$12$ keV band. 
When a 30\arcsec\ circle is used, the differential 
pn count rate is $(3.0 \pm 0.7)\times 10^{-3}$ cts s$^{-1}$ 
in the same energy band. 
Applying the same method to the MOS1 dataset, 
we obtain a differential count rate of 
$(8.6 \pm 2.8)\times 10^{-4}$ cts s$^{-1}$ 
in the $0.3$--$12$ keV band when we use a 12\arcsec\ 
extraction region, and $(6.9 \pm 3.8)\times 10^{-4}$ 
cts s$^{-1}$ for a 30\arcsec\ circle. 
For MOS2, the count rates are 
$(7.5 \pm 2.9)\times 10^{-4}$ cts s$^{-1}$ and 
$(7.5 \pm 3.8)\times 10^{-4}$ cts s$^{-1}$, respectively.
These values are consistent with the respective pn and MOS 
count rates determined from the 2002 dataset alone 
with the previous method, confirming that 
the contribution of SN~2002ap in 2003 is negligible. 

Taking the average of the three measurements for each detector, 
we obtain the pn, MOS1 and MOS2 count rates
listed in Table 1.  In the same table we have also reported 
the count rates in separate energy bands. The errors  
quoted in Table 1 are the errors in the mean from 
the three measurements (Gaussian propagation).

We point out that the analysis presented in this section supersedes 
the preliminary report on the X-ray colors and luminosity of SN~2002ap 
published in Sect. 4.2 of Soria \& Kong (2002). In that paper, we said that 
the source was very hard: in fact, this was caused by contamination 
from the nearby hard source CXOU J013623.5+154458, which has been 
properly subtracted here with the help of the Chandra datasets.

\subsection{Flux and luminosity of SN~2002ap}

The low signal-to-noise ratio of the pn spectrum 
does not allow a meaningful model fitting 
in {\footnotesize{XSPEC}}. However, we can at least constrain the spectrum 
of SN~2002ap by comparing its soft and hard X-ray colours  
with the colours expected for some simple spectral models.
The count rates inferred for the three pn narrow bands 
on 2002 February 2 (Table 1) are indicative of a soft spectrum. 
The conversion from count rates to emitted fluxes 
is strongly dependent on the absorbing column density. 
The foreground Galactic H{\footnotesize{\,I}} column density 
in the direction of M\,74 has been estimated 
as $4.8 \times 10^{20}$ cm$^{-2}$ 
(Dickey \& Lockman 1990). 
From the interstellar dust maps of Schlegel et al.~(1998), 
a reddening $E(B-V)= 0.072 \pm 0.012$ is obtained at the position 
of SN~2002ap. Using the empirical relation $n_{\rm H} = 1.79 \times 10^{21} 
A_{V} = 5.73 \times 10^{21} E(B-V)$ cm$^{-2}$ (Predehl \& Schmitt 1995), 
where $A_{V}$ is the extinction in the $V$ band, 
a Galactic column density $n_{\rm H} = 4.2^{+0.6}_{-0.7} 
\times 10^{20}$ cm$^{-2}$ is derived. To this value, we need to add 
the intrinsic absorption due to the circumstellar material. 
For this component, optical spectroscopic analysis 
of the Na{\footnotesize{\,I}} D lines shows 
that $n_{\rm H} = (1.15\pm0.05) \times 10^{20}$ cm$^{-2}$ 
(Takada-Hidai et al. 2002).
Hence, we can take $5 \times 10^{20}$ cm$^{-2}$ 
as a lower limit on the total absorbing column 
density. However, the X-ray spectrum may be 
more absorbed than the optical emission, hence 
we cannot obtain an upper limit for $n_{\rm H}$ 
from the optical observations. We shall discuss 
this issue in Section 3.1.

We then assumed that the spectrum could be approximated 
by one or two optically-thin thermal-plasma components 
(Raymond-Smith models in {\footnotesize{XSPEC}}, with solar abundance), 
and we investigated the range of temperatures and 
column densities consistent with the observed colours. 
We found (Fig.~1) that the spectrum is inconsistent 
with single-temperature models at $kT \simgt 1$ keV.
The observed X-ray colours are consistent 
for example with a single-temperature model with 
$kT \approx 0.5$ keV and $n_{\rm H} 
\approx 5 \times 10^{21}$ cm$^{-2}$; or with 
$kT \approx 0.75$ keV and $n_{\rm H} 
\approx 3 \times 10^{21}$ cm$^{-2}$. 
If we take the lowest allowed value of $n_{\rm H} 
\approx 5 \times 10^{20}$ cm$^{-2}$, 
the observed colours imply $kT \approx 0.85$ keV.
We cannot rule out two-temperature models, 
for example, with  
$kT_1 \approx 0.5$ keV, $kT_2 \approx 2$ keV, 
$n_{\rm H} \approx 1 \times 10^{21}$ cm$^{-2}$, 
where we have normalized the two components 
so that each contributes half of the emitted flux in 
the $0.3$--$12$ keV band. However, two-temperature models 
including a hot component with $kT \simgt 5$ keV 
are inconsistent with the soft colours observed 
for this source.

We calculated the fluxes 
for some representative single-temperature models 
consistent with the observed colours (Table 2). 
As noted earlier, most of the X-ray flux  
is detected below 2 keV. We estimate an emitted luminosity 
in the $0.3$--$2$ keV band ranging from a few times $10^{37}$ 
to $\approx 10^{38}$ erg s$^{-1}$ (depending on the spectral 
model), and an emitted 
luminosity in the $2$--$12$ keV band of 
$\approx 1.2 \times 10^{36}$ erg s$^{-1}$ for 
all spectral models. However, given the large 
uncertainties in the background subtraction, 
we can take $\approx 1 \times 10^{37}$ erg s$^{-1}$ 
as a safe upper limit for the emitted luminosity 
in the hard band. A distance of 7.3 Mpc has been assumed.

\begin{table*}
\caption{\xmm\ count rates for SN~2002ap on 2002 February 2, in units 
of $10^{-4}$ cts s$^{-1}$. 
The rates have been corrected for the telescope vignetting 
and the finite size of the extraction region. The values listed here are the averages 
of the results obtained with the three measurements described in Section 2.2.}
\label{crates} 
\centering 
         \begin{tabular}[width=textwidth]{lcccc}
            \hline
            \hline
            \noalign{\smallskip}
	Instrument & Count rate & Count rate & Count rate & Count rate \\
        & ($0.3$--$12.0$) keV & ($0.3$--$1$) keV & ($1$--$2$) keV & ($2$--$12$) keV\\
            \noalign{\smallskip}
            \hline\\
            \noalign{\smallskip} 
EPIC pn & $27.8\pm3.4$ & $19.5\pm2.6$ & $8.2\pm1.6$ & $0.3\pm2.0$ \\
EPIC MOS1 & $7.5\pm1.8$  & & & \\
EPIC MOS2 & $7.6\pm1.9$  & & & \\[5pt]
            \hline
         \end{tabular}
   \end{table*}

                            
\begin{figure}[t] 
\begin{center} 
\epsfig{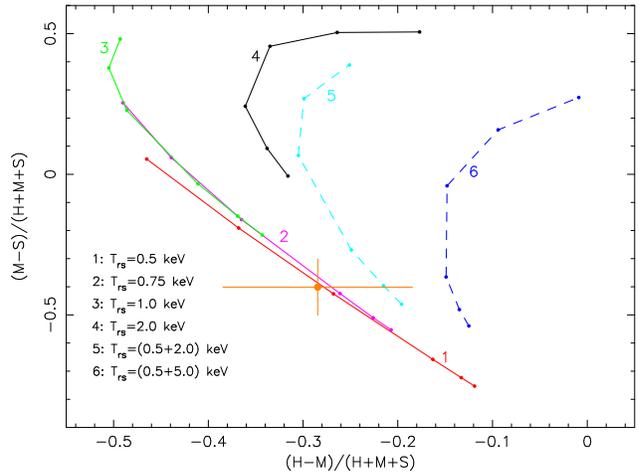}
\end{center}
\caption{Observed colours of SN~2002ap on 2002 February 2. 
Here $S$ is the pn count rate in the $0.3$--$1$ keV band, 
$M$ is the rate in the $1$--$2$ keV band, and $H$ the rate 
in the $2$--$12$ keV band (as listed in Table 1). The curves 
show the expected colours for some simple one- and two-temperature 
Raymond-Smith spectral models, at varying column densities. 
Along each curve, we plotted the colours corresponding 
to $n_{\rm H} = (0.5, 1.0, 2.0, 5.0, 7.5, 10) \times 10^{21}$ 
cm$^{-2}$ (column density increasing from the bottom to the top 
along each curve).}
\label{fig:sncolors}
\end{figure}


\begin{table*}
\caption{Observed and emitted fluxes  
of SN~2002ap on 2002 February 2, for three different spectral models 
consistent with the observed X-ray count rates in the EPIC pn and MOS 
bands (Table 1). A distance of 7.3 Mpc has been assumed. 
Fluxes are in units of $10^{-15}$ erg cm$^{-2}$ s$^{-1}$.} 
\label{fluxes} 
\centering 
         \begin{tabular}[width=textwidth]{lcccc}
            \hline
            \hline
            \noalign{\smallskip}
	Instrument & Flux & Flux  & Flux & Flux \\
        & ($0.3$--$12.0$) keV & ($0.3$--$1$) keV & ($1$--$2$) keV & ($2$--$12$) keV\\
            \noalign{\smallskip}
            \hline\\
& \multicolumn{4}{c}{model: $n_{\rm H} = 5 \times 10^{21}$ cm$^{-2}$; 
 $kT_{\rm {rs}} = 0.50$ keV; $Z=Z_{\odot}$}\\[5pt]  
\hline \\
            \noalign{\smallskip} 
pn observed flux & $4.9\pm0.6$ & $2.8\pm0.4$ & $1.9\pm0.4$ & $0.2\pm1.2$ \\
MOS1 observed flux & $4.0\pm1.0$  & & & \\
MOS2 observed flux & $4.1\pm1.0$  & & & \\[5pt]
pn emitted flux & $25.7\pm3.1$ & $21.5\pm2.9$ & $3.9\pm0.8$ & $0.2\pm1.2$ \\
MOS1 emitted flux & $21.1\pm5.1$  & & & \\
MOS2 emitted flux & $21.4\pm5.2$  & & & \\[5pt]
  \noalign{\smallskip}
            \hline \\
& \multicolumn{4}{c}{model: $n_{\rm H} = 3 \times 10^{21}$ cm$^{-2}$; 
 $kT_{\rm {rs}} = 0.75$ keV; $Z=Z_{\odot}$}\\[5pt]  
\hline \\
            \noalign{\smallskip} 
pn observed flux & $5.2\pm0.6$ & $2.9\pm0.4$ & $2.0\pm0.4$ & $0.2\pm1.2$ \\
MOS1 observed flux & $4.1\pm1.1$  & & & \\
MOS2 observed flux & $4.2\pm1.1$  & & & \\[5pt]
pn emitted flux & $12.4\pm1.5$ & $9.2\pm1.2$ & $3.1\pm0.7$ & $0.2\pm1.2$ \\
MOS1 emitted flux & $9.9\pm2.4$  & & & \\
MOS2 emitted flux & $10.0\pm2.4$  & & & \\[5pt]
  \noalign{\smallskip}
            \hline \\
& \multicolumn{4}{c}{model: $n_{\rm H} = 0.5 \times 10^{21}$ cm$^{-2}$; 
 $kT_{\rm {rs}} = 0.85$ keV; $Z=Z_{\odot}$}\\[5pt]  
\hline \\
            \noalign{\smallskip} 
pn observed flux & $5.0\pm0.6$ & $2.8\pm0.4$ & $2.0\pm0.4$ & $0.2\pm1.2$ \\
MOS1 observed flux & $4.1\pm1.1$  & & & \\
MOS2 observed flux & $4.2\pm1.1$  & & & \\[5pt]
pn emitted flux & $5.9\pm0.7$ & $3.6\pm0.5$ & $2.0\pm0.4$ & $0.2\pm1.2$ \\
MOS1 emitted flux & $4.9\pm1.3$  & & & \\
MOS2 emitted flux & $5.0\pm1.3$  & & & \\[5pt]
\hline\\
	\end{tabular}
   \end{table*}

\subsection{Two ``ultra-luminous'' sources?}
 
Two variable X-ray sources were detected by \xmm\ with 
peak emitted luminosities $\simgt 10^{39}$\lum.   
One of them, XMMU J013636.5$+$155036, was detected only 
in the 2002 February EPIC observation (Soria \& Kong 2002). 
It was not found in the 2003 January co-added EPIC image, from 
which we estimate that it was at least 30 times fainter 
than 11 months before, in the $0.3$--$12$ keV band. 
It was also undetected in both {\it Chandra} 
images from 2001. We refined the spectral analysis 
presented in Soria \& Kong (2002) by co-adding 
the pn and MOS data (thus increasing the significance of possible 
line features), with a program written by M. Page, 
and by using updated response matrices 
and more recent calibration files in the {\footnotesize{SAS}}. 
The combined EPIC spectrum from 2002 February 
is well fitted ($\chi^2_\nu = 30.9/36$) 
by a simple power law with $\Gamma \approx 1.9$, 
and total (Galactic plus intrinsic) absorption 
$n_{\rm H} \approx 1.8 \times 10^{21}$ cm$^{-2}$ 
(Table 3). This corresponds to an emitted 
luminosity $\approx 1.6 \times 10^{39}$\lum\ in the 
$0.3$--$12$ keV band. Adding a blackbody or multicolor disk-blackbody 
component at $kT \approx 0.15$ keV does not improve the fit 
significantly ($\chi^2_\nu = 27.8/34$). 
An absorbed multicolor disk-blackbody model does not provide a good fit 
(best-fit $kT_{\rm in} = 1.2 \pm 0.2$ keV, but $\chi^2_\nu = 47.9/36$).
We also analysed the lightcurve but found no periodicity in the 1--10,000 s range. 
The count rate is consistent with the source being constant 
over the duration of the 2002 \xmm\ observation. 
Its emitted luminosity in 2003 
must be $< 5 \times 10^{37}$ erg s$^{-1}$ in the 
$0.3$--$12$ keV band. The upper limit, at the 3-$\sigma$ significance 
level, was estimated with the assumption of a $\Gamma = 1.9$ power-law 
spectrum and total $n_{\rm H} = 1.8 \times 10^{21}$ cm$^{-2}$.

The other bright source in the M\,74 field, CXOU J13651.1$+$154547, 
is detected in both {\it Chandra} and 
both \xmm\ observations\footnote{This source was not
discussed in Soria \& Kong (2002) because in that paper 
we focussed only on transient sources newly detected by XMM}. 
In the first three 
observations it exhibited strong variability over timescales 
of a few thousand seconds, with flux changes as large as 
an order of magnitude (Krauss et al.~2003). If we assume that 
this source belongs to M\,74 and is not a background AGN\footnote{Two 
candidate ``ultra-luminous'' sources in the field of nearby galaxies 
have recently been recognised as background AGN 
(Masetti et al.~2003; Foschini et al.~2002). The possibility that 
CXOU J13651.1$+$154547 is a background object was discussed 
by Krauss et al.~(2003), but considered unlikely 
in the absence of any radio or optical counterpart.},
typical luminosities inferred from the 2001--2002 data 
were $\approx 5 \times 10^{38}$ 
\lum\ in the faint state, with peaks of up to $\approx 8 \times 10^{39}$ 
during the flares. Variability over shorter timescales ($\sim 100$ s)
was also observed. Spectral analysis showed that the X-ray emission 
was harder when the source was brighter (Krauss et al.~2003).

Data analysis for the 2003 \xmm\ observation 
is complicated by the location of the source on a CCD gap in the pn. 
We eliminated the contamination of spurious events 
along the chip edge by extracting only "flag-0" \& "pattern-0" 
events. Fortunately, the source is located 
sufficiently far away from any chip gaps in the MOS.
Taking into account vignetting (the source is 5\arcmin\ off-axis) 
and the fraction of pn events lost in the chip gap, 
we obtained the combined EPIC lightcurve shown 
in Fig.~2.
The observed count rate appears to be much less variable than 
in 2001 and 2002, with no signs of strong flares. 
We carried out Fourier analysis of the data 
but found no significant periodicity in the 1--10,000 s range.

The co-added pn and MOS spectrum can be fitted in the $0.3$--$12$ keV band  
with a simple power-law with $\Gamma \approx 2.5\pm0.5$, absorbed by 
a total column density 
$n_{\rm H} \approx 1.7 \times 10^{21}$ cm$^{-2}$ 
($\chi^2_{\nu} = 1.16$ for 18 d.o.f; Fig.~3 and Table 4). 
Adding a multicolour black-body component or using comptonized 
black-body models (eg, comptt, bmc or thcompds in {\footnotesize{XSPEC}}) 
does not significantly improve the fit. When the Comptonization 
model bmc\footnote{The bmc model (Shrader \& Titarchuk 1999) 
describes the thermal or bulk motion 
Comptonization of a blackbody seed-photon component.} is used, 
the temperature of the seed thermal component 
is constrained to be $\simlt 0.26$ keV (90\% confidence level).
Using the Comptonization 
model comptt\footnote{comptt (Titarchuk 1994) 
approximates the seed-photon input spectrum 
with the Wien tail of the blackbody spectrum. Thus, it is simpler but less 
accurate than the bmc model.}, the seed photon temperature is $\simlt 0.23$ keV.
Finally, we checked that the observed spectrum is not consistent 
with an absorbed multicolor disk-blackbody ($\chi^2_{\nu} = 1.56$ for 18 d.o.f).
Using the power-law spectral model, we obtain an emitted flux 
$1.3 \times 10^{-13}$ erg cm$^{-2}$ s$^{-1}$ in
the $0.3$--$12$ keV band; for the bmc model, the flux 
is $1.2 \times 10^{-13}$ erg cm$^{-2}$ s$^{-1}$. 
At the assumed distance of M\,74, the emitted luminosity is 
$\approx 8 \times 10^{38}$ erg s$^{-1}$ for both spectral models.

As an aside, we note that XMMU J013627.2$+$155005 
is another bright transient source 
seen in 2002 (when it had a hard power-law spectrum 
and a luminosity $\approx 3 \times 10^{38}$ 
erg s$^{-1}$, Soria \& Kong 2002) but not in 2003, 
nor in the previous {\it Chandra} observations.

                            
\begin{figure}[t] 
\begin{center} 
\epsfig{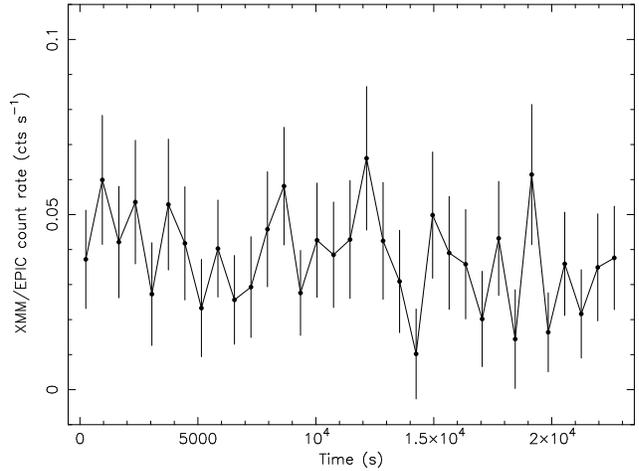}
\end{center}
\caption{Lightcurve of CXOU J13651.1$+$154547 for the 
\xmm\ 2003 January observation. The data have been binned
into 700 s intervals. The count rate plotted here 
is the total rate that would 
have been detected by the three EPIC instruments combined, 
in the $0.3$--$12$ keV band, if the source 
had been observed on-axis.}
\label{fig:ulxlc}
\end{figure}


                            
\begin{figure}[t] 
\begin{center} 
\epsfig{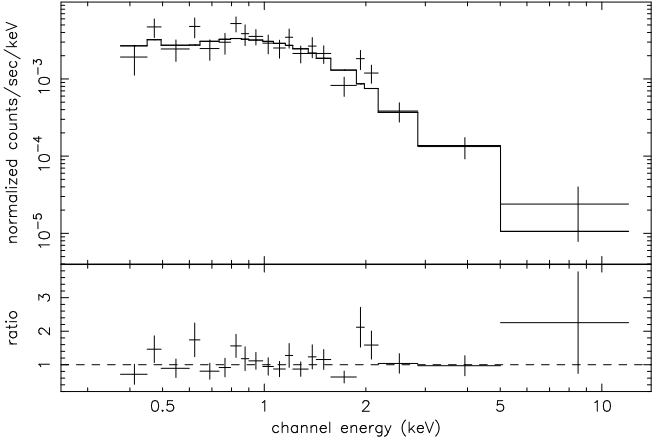}
\end{center}
\caption{Spectrum of CXOU J13651.1$+$154547 for the 
\xmm\ 2003 January observation. The pn and MOS spectral data 
have been re-sampled and co-added to produce a combined EPIC 
spectrum. The background-subtracted spectrum has then been grouped  
to achieve a signal-to-noise ratio $> 4$ for each bin. 
The fitted model is an absorbed power-law of index $\Gamma \approx 2.5$.}
\label{fig:ulxspec}
\end{figure}


\begin{table}[t]
\caption{
   {\small XSPEC} best-fit parameters 
   for the X-ray spectrum of XMMU J013636.5$+$155036 
	in 2002 February, 
	from the combined \xmm\ EPIC dataset. The foreground 
	line-of-sight absorption column density 
	$n_{\rm {H,\,Gal}}$ has been fixed at $0.5 \times 10^{21}$~cm$^{-2}$. 
	See http://heasarc.gsfc.nasa.gov/docs/xanadu/xspec  
	for a full description of the XSPEC spectral models.}
\centering 
\begin{tabular}{@{}lrr} 
\hline
\hline 
   \noalign{\smallskip}
\multicolumn{3}{c}{model: wabs$_{\rm Gal}$ $\times$ wabs $\times$ power-law  }\\[5pt]  
\hline \\ 
$n_{\rm H}~(\times 10^{21}$~cm$^{-2})$  & $1.3^{+0.6}_{-0.5}$   \\[5pt] 
$\Gamma$       & $1.91^{+0.21}_{-0.17}$ \\[5pt]
$K_{\rm pl}~(\times 10^{-5})$    & $4.0^{+0.9}_{-0.7}$ \\ [5pt]   
\hline \\ 
$\chi_\nu^2$~(dof) &  0.86~(36)  \\ [5pt] 
$L_{\rm{0.3-12}}~(\times 10^{39}$~erg~s$^{-1})$ & $1.6^{+0.1}_{-0.3}$  \\[5pt]
\hline \\
\end{tabular}     
\end{table}

\begin{table}
\caption{
   {\small XSPEC} best-fit parameters 
   for CXOU J13651.1$+$154547 in 2003 January, 
	from the combined \xmm\ EPIC dataset. 
	$n_{\rm {H,\,Gal}} \equiv 0.5 \times 10^{21}$~cm$^{-2}$ as in Table 3.}
\centering 
\begin{tabular}{@{}lrr} 
\hline
\hline 
   \noalign{\smallskip}
\multicolumn{3}{c}{model: wabs$_{\rm Gal}$ $\times$ wabs $\times$ power-law  }\\[5pt]  
\hline \\ 
$n_{\rm H}~(\times 10^{21}$~cm$^{-2})$  & $1.2^{+0.9}_{-0.8}$   \\[5pt] 
$\Gamma$       & $2.46^{+0.47}_{-0.41}$ \\[5pt]
$K_{\rm pl}~(\times 10^{-5})$    & $2.5^{+1.1}_{-0.7}$ \\ [5pt]   
\hline \\ 
$\chi_\nu^2$~(dof) &  1.16~(18)  \\ [5pt] 
$L_{\rm{0.3-12}}~(\times 10^{38}$~erg~s$^{-1})$ & $8.0^{+2.8}_{-3.8}$  \\[5pt]
\hline 
\hline \\
\multicolumn{3}{c}{model: wabs$_{\rm Gal}$ $\times$ wabs $\times$ bmc  }\\[5pt]  
\hline \\ 
$n_{\rm H}~(\times 10^{21}$~cm$^{-2})$  & $1.2^{+4.3}_{-1.2}$   \\[5pt] 
$T_{\rm bb}$~(keV) & $0.13^{+0.13}_{-0.13}$ \\[5pt] 
$\Gamma$       & $2.22^{+0.82}_{-1.22}$ \\[5pt]
$K_{\rm bmc}~(\times 10^{-7})$    & $9.2^{+5.0}_{-5.0}$ \\ [5pt]   
\hline \\ 
$\chi_\nu^2$~(dof) &  1.25~(16)  \\ [5pt] 
$L_{\rm{0.3-12}}~(\times 10^{38}$~erg~s$^{-1})$ & $7.7^{+1.2}_{-3.0}$  \\[5pt]
\hline \\
\end{tabular}     
\end{table}

\subsection{Colour and luminosity distribution of the other X-ray sources}

In addition to an individual study of SN~2002ap, XMMU J013636.5$+$155036 
and CXOU J013651.1$+$154547, we investigated the color and luminosity 
distribution of the discrete 
source population in M\,74, to distinguish different physical 
classes of sources and estimate the relative fraction of 
X-ray binaries (XRBs) and supernova remnants (SNRs). 
The spatial resolution of the \xmm\ EPIC cameras is not sufficient 
to resolve many faint sources in the inner disk; therefore, we used 
the co-added 2001 June $+$ October {\it Chandra} ACIS-S 
observations for this study, achieving a detection limit 
of $\approx 10^{-4}$ ACIS-S cts s$^{-1}$.

Seventy-four X-ray sources (not including two obvious 
foreground stars) are detected with {\it wavdetect} 
at $> 3.5\sigma$ significance inside the $D_{25}$ ellipse 
of M\,74, in the {\it Chandra} ACIS-S image (Appendix A.1). 
The cumulative count rate distribution (Fig.~4)
is fitted by a simple power-law of index $\alpha \approx -1.0$, 
for the 46 sources detected with an ACIS count rate $> 3 \times 10^{-4}$ 
cts s$^{-1}$, which we take as the completeness limit. 
Assuming $n_{\rm H} = 1 \times 10^{21}$ cm$^{-2}$, 
and a power-law spectrum with $\Gamma = 1.7$, this corresponds 
to an emitted luminosity $\approx 2 \times 10^{37}$\lum.
Using the results of the {\it Chandra} 1Ms Deep Field South exposure 
(Rosati et al.~2002), we estimate that as many as 20 of these 46 
sources may be background AGN. We also estimate that 3 to 5 of the 11 
sources detected with an ACIS count rate $> 10^{-3}$, and 
at most 1 of the 5 brightest sources (count rate $> 3 \times 10^{-3}$) 
may be from the background.
After removing the estimated fraction of AGN at various fluxes, 
we obtain that the true slope of the cumulative count rate distribution 
is $\alpha \approx -0.9$ ($\alpha \approx -0.8$ for sources 
more luminous than $\approx 10^{38}$\lum; this is in agreement 
with the slope found by Soria \& Kong (2002)).

A power-law luminosity function is consistent with the distribution 
observed in disks of other spiral galaxies with moderately active 
star formation (eg, $\alpha \approx -0.8$ in M\,101, see Pence et al.~2001; 
$\alpha \approx -1.1$--\,$-0.9$ in the disk of M\,31, see Kong et al.~2003). 
A steeper power-law index ($\alpha \approx -1.7$) is generally observed in spiral 
bulges, which are dominated by old stellar populations, while flatter 
slopes ($\alpha \approx -0.5$) are typical of starburst galaxies 
(eg, the Antennae and M\,82; see Zezas \& Fabbiano 2002 and references 
therein).

Colour-colour plots (Figs.~4 and 5) are an effective way 
of separating the discrete sources into separate physical groups: 
soft and hard XRBs, SNRs 
and supersoft sources. We chose the colour indices 
[$(H-M)/(H+M+S)$,\,$(M-S)/(H+M+S)$], following 
Prestwich et al.\ (2003).
The spectral models overplotted in the diagrams 
are: power-laws with photon indices 
$\Gamma = 1.3$, $\Gamma = 1.7$ and $\Gamma = 2.0$ (characteristic 
of XRBs in the hard state); disk-blackbody 
with inner-disk temperatures 
$kT_{\rm in} = 0.5$ keV and $kT_{\rm in} = 1.0$ keV (typical 
of XRBs in the soft state);
optically-thin, single-temperature thermal plasma at $kT_{\rm rs} = 0.5$ keV 
(typical of SNRs); blackbody at $kT_{\rm bb} = 0.1$ keV (typical 
of supersoft sources).
Along each model line, the column density increases 
from the bottom to the top (models 1 and 2) or clockwise (models 3--7): 
from $n_{\rm H} = 5 \times 10^{20}$~cm$^{-2}$ 
(line-of-sight foreground absorption for M\,74) 
to $n_{\rm H} =  5 \times 10^{21}$~cm$^{-2}$ 
for the blackbody model, 
to $n_{\rm H} =  7.5 \times 10^{21}$~cm$^{-2}$ 
for the optically-thin thermal plasma model, 
to $n_{\rm H} =  2 \times 10^{22}$~cm$^{-2}$ 
for the disk-blackbody models, and 
to $n_{\rm H} =  5 \times 10^{22}$~cm$^{-2}$ 
for the power-law models. In calculating the expected colours 
for each of those spectral models, we took into account 
the time-dependent degradation of the ACIS detector at soft channel energies. 

We also plotted in Figs.~5 and 6 the approximate colours 
that SN~2002ap would have had if it had been observed 
with the ACIS-S detector\footnote{The conversion 
between the \xmm\ and {\it Chandra} count rates in the three bands 
depends of course on the assumed spectral model and absorption, 
but these variations are small for the range of column densities 
and temperatures considered in Table 2. We converted the colours 
using the correct response matrices generated by the {\footnotesize{CIAO}} 
and {\footnotesize{SAS}} software, rather than the default 
(pre-launch) responses from {\footnotesize{PIMMS}}.}. 
The source colours fall in the region of the diagam which in star-forming 
galaxies is usually populated by thermal SNRs (eg, Prestwich et al.~2003; 
Soria \& Wu 2003; Soria \& Kong 2003).

                            
\begin{figure}[t] 
\begin{center} 
\epsfig{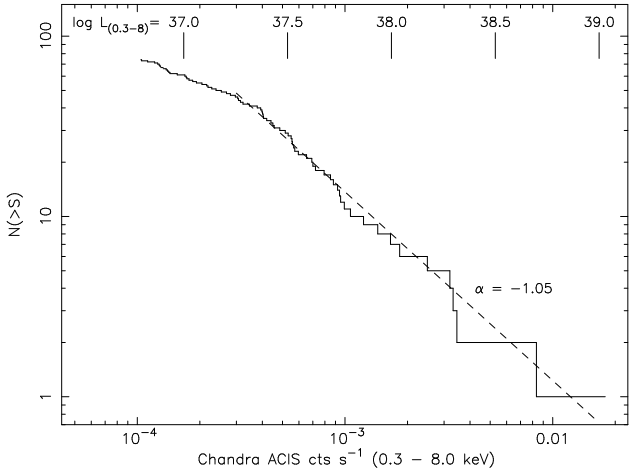}
\end{center}
\caption{Cumulative count-rate distribution ("luminosity 
function") for the 74 sources found inside the $D_{25}$ ellipse 
in the combined {\it Chandra} observations. For each source, 
the count rate plotted here is the average of 
the 2001 June and October count rates. Two obvious 
foreground stars are not included. See Sect.~2.5 for an estimate 
of the background AGN contribution.}
\label{fig:chandralum}
\end{figure}


                            
\begin{figure}[t] 
\begin{center} 
\epsfig{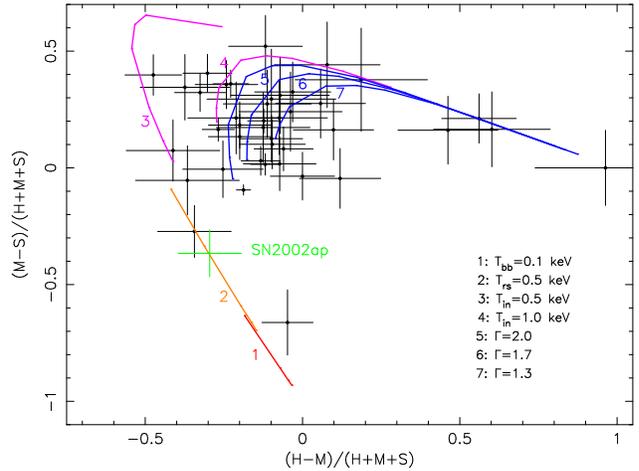}
\end{center}
\caption{
X-ray colour-colour plot of the 46 brightest sources (those with 
an average count rate $>3 \times 10^{-4}$ cts s$^{-1}$)
found inside the $D_{25}$ ellipse of M\,74 in the two {\it Chandra}  
observations. 
The three bands are: $S= 0.3$--$1.0$ keV; 
$M= 1.0$--$2.0$ keV; $H= 2.0$--$8.0$ keV. 
We overplotted the expected colours for some 
basic spectral models, which help us separate the different 
physical classes of sources. Along each model curve, 
the absorbing column density increases from the bottom to top 
(models 1--2) or clockwise (models 3--7). 
For comparison, we also plotted the approximate colour 
that SN~2002ap would have had in 2002 February 
if it had been observed by {\it Chandra} ACIS-S3.}
\label{fig:chandracol1}
\end{figure}


                            
\begin{figure} 
\begin{center} 
\epsfig{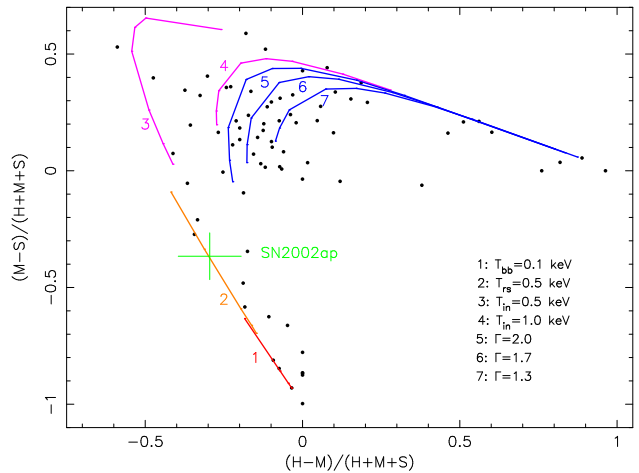}
\end{center}
\caption{
X-ray colour-colour plot of all the sources 
found inside the $D_{25}$ ellipse of M\,74 in the two {\it Chandra} 
observations, without error bars. 
The three energy bands and the spectral 
models have been chosen like in Fig.~5.}
\label{fig:chandracol2}
\end{figure}


\section{Discussion}

\subsection{The X-ray behaviour of SN~2002ap}

Type Ib/c SNe are rare occurrences in nearby galaxies, and they are only
rarely observed in the X-rays.  It is not clear whether there is a
``typical" X-ray behaviour. In fact, the two best-monitored previous cases
of Type Ic X-ray SNe, 1994I and 1998bw, have substantially different X-ray
light curves at early times (Fig.~7). SN~1998bw exhibited early 
X-ray emission (detected by {\it BeppoSAX}) in the $0.1$--$10$ keV range, 
starting 10 hours after the explosion (Pian et al. 2000). 
The initial decline was slow: only a factor $\approx 2.5$ 
over the first 6 months. An \xmm\ measurement nearly 4 years later 
shows that the X-ray decline must have been faster at later epochs (Fig.~7; 
Pian et al. 2003). In contrast, SN~1994I was not
detected by {\it ROSAT} until day 82 after the explosion (Immler et al. 1998).  
However, its late-epoch temporal behaviour (i.e. after day 82), sampled
by {\it Chandra} (Immler et al. 2002), is consistent with that of SN~1998bw (Fig.~7).

\begin{figure}[t] 
\hspace{-0.4cm}
\epsfig{figure=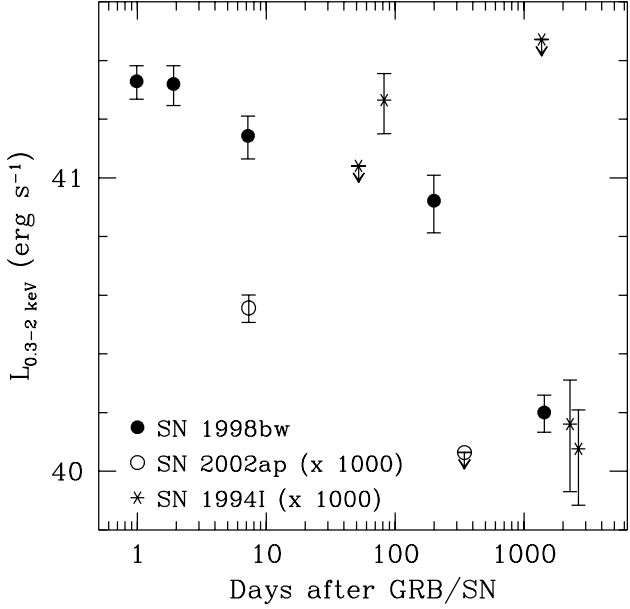,width=9.6cm}
\caption{The lightcurves of Type Ic SNe 1998bw, 1994I and 2002ap in the
$0.3$--$2$ keV band.  The datapoints for SN~1998bw and SN~1994I are taken 
from Pian et al. (2003) and references therein, and from Immler et al. (2002),
respectively.} 
\label{fig:sn3}
\end{figure}

The temporal behaviour of SN~2002ap is overall more similar to that of
SN~1998bw, in the sense that both SNe showed prompt X-ray emission. 
The 2003 upper limit on the X-ray flux of SN~2002ap 
is at least 7 times fainter than the 2002 detection: this is roughly
consistent with the decay rate of SN~1998bw (Fig.~7).  
The early X-ray detection of SN~1998bw may suggest the presence of a jet, 
in which prompt, non-thermal radiation can be efficiently produced, 
similar to GRB afterglow radiation (see Pian et al. 2003).
However, the absence of hard X-ray emission in SN~2002ap (the upper limit on
the $2$--$12$ keV flux implies a luminosity at least a factor $\sim$1000 lower
than in SN~1998bw) argues against the presence of highly relativistic
conditions at the forward (circumstellar) shock. 
This SN was also less radio-luminous than SN~1998bw (Berger et al. 2002).  
The lower energy emitted in the X-ray and
radio bands is consistent with the estimate of the total energy of SN~2002ap
($5 \times 10^{51}$ erg s$^{-1}$), which is an order of magnitude smaller
than in SN~1998bw (Mazzali et al.~2002).
While the consistency of the temporal X-ray behaviour of SN~2002ap with that
of SN~1998bw could support its classification as a hypernova, suggested on
the basis of optical observations (Mazzali et al.~2002; Foley et al.~2003), 
the lower total energy may be insufficient for the
development of a strongly asymmetric explosion, as required for the
formation of a GRB. Indeed, a search for gamma-ray signal during the week
prior to the SN~2002ap explosion resulted in no detection (Hurley et al.~2002).

X-ray emission from a SN may originate either from the forward
shock or from the reverse shock (eg, Immler \& Lewin 2002
and Chevalier \& Fransson 2002 for reviews). Following the self-similar
analytical approximation of Fransson et al.~(1996), 
and of Chevalier \& Fransson (1994), one can
parameterise the radial density profile $\rho$ of the ejecta 
as a function of time $t$, as $\rho = K t^{-3} (r/t)^{-n}$, 
where: $r$ is the radial distance from the centre 
of the progenitor; the proportionality constant $K$ depends on the velocity 
$V$ of the ejecta at the reverse shock (maximum 
velocity of the ejecta at any given time); the choice 
of the power-law index $n$ depends on the type of progenitor star.
The circumstellar wind density $\rho_{\rm w}$ can
be parameterised as: \begin{equation} \rho_{\rm w} = \frac{\dot{M}}{4\pi
v_{\rm w} r^2_0} \, \left(\frac{r_0}{r}\right)^s \end{equation} 
where $r_0$ is a reference radius corresponding 
to the mass-loss rate $\dot{M}/v_{\rm w}$ ($r_0 = 10^{15}$ cm in 
Fransson et al.~1996), and $1.5 \simlt s \simlt 2$ 
(Fransson et al.~1996).  
The mass-loss rate $\dot{M} \approx 5 \times 10^{-7} M_{\odot}$
yr$^{-1}$ for the progenitor of SN~2002ap (Berger et al. 2002).  
The stellar-wind velocity $v_{\rm w}$ is generally of the same order as the escape 
velocity from the progenitor star, that is $\sim 10^3$ km s$^{-1}$ 
in the case of Type Ic SN progenitors.
As noted earlier, the
low radio and hard X-ray luminosities of SN~2002ap seem to rule out 
a relativistic speed for the ejecta: it is more likely 
that $V \sim (1$--$3) \times 10^4$ km s$^{-1}$, as is the case 
in most core-collapse SNe (eg, Matzner \& McKee 1999).

In the thin shell approximation, and assuming electron-ion 
equipartition, the forward shock has a temperature 
\begin{equation} 
kT_{\rm f} \simeq 196 \, \mu_{\rm s} \, \left(\frac{n-3}{n-s}\right)^2 
\, V_4^2 \ \ {\rm keV} \approx 150 \, V_4^2 \ \ {\rm keV},
\end{equation}
where $V_4 \equiv V/(10^4 \rm{\,km~s}^{-1})$,  
and $\mu_{\rm s}$ is the mean mass per particle. 
$\mu_{\rm s}$ is a function of chemical abundance: $\mu_{\rm s} = 0.61$ 
for solar abundance, and, more generally, $0.5 < \mu_{\rm s} \simlt 1$.
We have also assumed $n \gg 3$, $n \gg s$ (Fransson et al.~1996).
The temperature of the reverse shock 
is (Fransson et al.~1996): 
\begin{equation}
kT_{\rm r} = \left(\frac{3-s}{n-3}\right)^2 \,kT_{\rm f} 
= 196 \, \mu_{\rm s} \left(\frac{3-s}{n-s}\right)^2 \, V_4^2 \ \ {\rm keV}.
\end{equation}

The observed colours of SN~2002ap imply temperatures 
$0.5 \simlt kT \simlt 0.9$ keV, suggesting 
that the emission comes from the reverse shock, and that 
the density gradient of the ejecta is very steep, 
with $n \simgt 20$, from Eq.(3). 

The cooling timescale $t_{\rm cool}$ 
of the reverse shock, at each time $t$ 
after the SN event, depends on $n$, $s$, $V$, $\dot{M}$ 
and $v_{\rm w}$ (Equation (3.14) in Fransson et al.~1996). 
Taking $s = 2$, and $t = 5$ d (time of the {\it XMM-Newton} 
observation), one obtains: 
\begin{equation}
t_{\rm cool} = 1.5 \times 10^7\, 
\frac{V_4^{5.34}}{(n-3)\,(n-4)\,(n-2)^{3.34}}\,
\frac{v_{{\rm w},3}}{\dot{M}_{-6}} \ \ \rm{d},
\end{equation}
where $\dot{M}_{-6}$ is the mass-loss rate in units of $10^{-6} M_{\odot}$ 
yr$^{-1}$, and $v_{{\rm w},3}$ the wind speed in units of $10^3$ km s$^{-1}$. 
For $n \simgt 20$, and $v_{{\rm w},3}/\dot{M}_{-6} 
\approx 1$, $t_{\rm cool} \approx 10^7 (V_4/n)^{5.34}$ d, i.e., only a 
few days or weeks. We would not expect the reverse shock 
to remain radiative after a year.

The total luminosity emitted by the reverse shock is (for $s=2$): 
\begin{equation}
L_{\rm r} = 1.6 \times 10^{38}\, 
V_4^3 \, \frac{(n-3)\,(n-4)}{(n-2)^3}\,
\frac{\dot{M}_{-6}}{v_{{\rm w},3}} \ \ \rm{erg~s}^{-1}.
\end{equation}
This simple approximation gives an emitted luminosity 
$\sim 10^{37}$ erg s$^{-1}$, in agreement with the observations 
(Table 2) if the absorbing column density is $\simlt 10^{21}$ cm$^{-2}$.
On the other hand, the same model predicts an X-ray luminosity from 
the forward shock $\sim 10^{31}$--$10^{32}$ erg~s$^{-1}$, 
entirely negligible (Fransson et al.~1996).
As the reverse shock cools down radiatively, 
a cold shell is formed between the reverse shock and the contact 
discontinuity, with a column density that scales 
$\sim (n-4)\,(Vt)^{-1}\,(\dot{M}/v_{{\rm w}})$ 
(Fransson et al.~1996). At $t = 5$ d after the event, 
and for values of the physical parameters suitable 
to SN 2002ap, this implies an intrinsic column density 
$\approx (1$--$2) \times 10^{20}$ cm$^{-2}$.

In conclusion, we argue that the low temperature 
($kT \approx 0.8$ keV) of the X-ray emission observed 
from SN 2002ap on 2002 February 2 points to a radiatively-cooling reverse
shock as the origin of the X-rays. If so, the fact 
that it was already detected 5 d after the event implies that 
the cool shell between the reverse shock and the contact 
discontinuity was already optically thin. 
The ratio of mass-loss rate over stellar-wind velocity 
is the crucial factor determining the column density 
across the cool shell (scaling as $(\dot{M}/v_{{\rm w}})$) 
and the luminosity of the forward shock (scaling as $(\dot{M}/v_{{\rm w}})^2$). 
For Type Ic SNe, which originate from a massive 
but more compact progenitor, this factor may be $\sim 10^3$ times 
smaller than for Type II SNe (eg, SN 1993J; SN 1999em). Hence, 
in the latter cases the cool shell is generally optically 
thick to the soft X-ray emission until a few weeks 
or months after the event (Fransson et al.~1996; Immler \& Lewin 2002), 
and only hard X-ray emission from the forward shock may be detected 
in the initial phases. In this scenario, the late X-ray detection of SN 1994I 
suggests a higher $\dot{M}/v_{{\rm w}}$ ratio for this object 
(Immler et al.~2002), than for SN 1998bw and SN 2002ap, despite 
all three being classified as Type Ic events.

The limited S/N and the scarce temporal information do not allow us to measure the X-ray
flux decline timescale in SN~2002ap, and therefore we are unable to investigate 
in more detail the density of the circumstellar medium as a function 
of radial distance from the SN, and the cooling timescale of the shocked ejecta. 
More cases of Type Ic SNe must be followed with \xmm, 
to understand the emission processes, constrain 
the parameters of the circumstellar medium and detect possible 
deviations from isotropy in the geometry of the explosion.

\subsection{Nature of the brightest X-ray sources}

Accreting X-ray sources are usually classified as ``ultra-luminous'' 
(ULXs) when they are persistently brighter than the "classical" 
Eddington limit of a $\approx 7 M_{\odot}$ BH 
(eg, Colbert \& Mushotzky 1999; Roberts \& Warwick 2000; Makishima et al.~2000).
In 2002 February, XMMU J013636.5$+$155036 was the brightest source 
in M\,74 with a luminosity $\approx 1.6 \times 10^{39}$\lum, 
approximately constant over the 25-ks observation, 
and a featureless power-law X-ray spectrum. Its location 
in a star-forming region suggested that the source was 
a high-mass X-ray binary. However, the source was no longer 
detected in 2003, its flux having declined by at least 
30 times, and we cannot determine  
how long the high state may have lasted. Repeated 
observations in the future will be necessary to determine 
its duty cycle.

The other candidate ULX in M\,74, CXOU J013651.1$+$154547, 
showed little variability in the 2003 January observation, 
at a luminosity of $\approx 8 \times 10^{38}$\lum. 
The persistent component of the X-ray emission 
was $\approx 5 \times 10^{38}$\lum\ in 2001 October 
and 2002 February (Krauss et al. 2003). Hence, 
the ``steady'' X-ray luminosity appears to be always 
less than the Eddington limit for a "canonical" $7 M_{\odot}$ BH. 
(This is still the case even for the larger distance 
to M\,74 assumed by Krauss et al. 2003).
The new observation suggests that the X-ray spectrum 
in the non-flaring state is marginally softer 
than during flares, and that it is dominated 
by the non-thermal (power-law) component, usually 
interpreted as Compton-scattered emission. 
Any additional blackbody or disk-blackbody components 
(required to provide the soft seed photons) 
are constrained to have temperatures $\simlt 0.25$ keV.
If confirmed, such a low temperature is consistent with 
a standard geometrically-thin accretion disk around 
an intermediate-mass BH (Makishima et al.~2000).
Alternatively, it would also be consistent with thermal emission 
from Compton-thick outflows (King \& Pounds 2003) 
which may occur for super-Eddington accretion rates 
($\dot{M}>\dot{M}_{\rm Edd}$). 

In the first three observations, hard X-ray flares were detected 
on top of the persistent component, on timescales of a few thousand 
seconds. During the flares, the source exceeded 
the $7 M_{\odot}$ Eddington limit by up to a factor 
of 10, but only for short periods of time: none of the flares 
lasted longer than $\approx 1$ h.
The rapid variability was interpreted (Krauss et al. 2003) 
as evidence of a beamed emission component, 
possibly associated with a jet. 
It may not be necessary to invoke a micro-quasar scenario 
to explain the flares: short episodes of super-Eddington 
emission are sometimes also found in neutron-star XRBs. 
For example, the high-mass XRB LMC X-4 
shows a persistent luminosity $\approx 2 \times 10^{38}$\lum\ 
$\approx L_{\rm{Edd}}$ for a neutron star, 
with repeated bursts/flares (typical duration $\sim 100$ s) 
reaching peak luminosities 
$\approx 2 \times 10^{39}$\lum\ (Moon et al. 2003).
The ``Rapid Burster'' MXB 1730$-$335 and GRO J1744$-$28 are 
two other examples of rapid, repeated flaring (Lewin et al.~1996).
The X-ray bursts in these systems are generally thought 
to be caused by the spasmodic release of gravitational 
potential energy (``Type-II'' bursts, as opposed 
to the thermo-nuclear ``Type-I'' bursts; eg, Lewin 1995). 
However, the exact mechanism is still unclear 
and may differ from source to source.
Thermal-viscous instabilities in the inner disk region 
have been invoked to explain the inhomogeneous accretion 
(Cannizzo 1996, 1997). Alternatively, the flares have been 
attributed to the accumulation and subsequent sudden 
release of accreting matter at a centrifugal barrier; 
in the case of a neutron star, the barrier 
can be created by the magnetosphere 
of the rapidly rotating compact object (Bann 1977, 1979). 
Mass transfer instabilities or inhomogeneous winds 
from the companion star have also been suggested 
(Vogt \& Penrod 1983).
Further X-ray studies of CXOU J013651.1$+$154547 
will be required to ascertain whether its flaring behaviour
could be an example of Type-II bursts in 
a stellar-mass BH XRB, and to determine 
what physical mechanism is responsible 
for the intermittent accretion, or for the intermittent 
ejections in the micro-quasar scenario.

\subsection{X-ray colours of the point sources}

The discrete X-ray sources are roughly clustered 
into three main groups in the colour-colour plots 
(Figs.~5 and 6). Firstly, ``classical'' supersoft sources have  
detectable emission only in the soft band and 
are therefore located at $(H-M)/(S+M+H) \approx 0$, $(M-S)/(S+M+H) \simlt -0.7$.
Secondly, a group of soft sources with detectable emission in the 
soft and medium band is located at   
$-0.7 \simlt (M-S)/(S+M+H) \simlt -0.2$. It has been 
suggested (Prestwich et al.\ 2003) that thermal 
SNRs dominate this class of sources. 
Thirdly, most XRBs have $-0.2 \simlt (M-S)/(S+M+H) \simlt 0.5$, with 
a broad spread in $(H-M)/(S+M+H)$.  
(See also Soria \& Wu (2003) and Soria \& Kong (2003) 
for a discussion of colour-colour plots in other nearby 
spiral galaxies, and Di\,Stefano \& Kong (2003) 
for a classification of soft and supersoft sources.) 
A preliminary comparison between the two {\it Chandra} 
and two \xmm\ observations 
shows that at least three bright ($\sim 10^{38}$\lum) 
candidate XRBs exhibit spectral transitions between 
a harder and a softer state. This is beyond the scope 
of this paper and is left to further work.

Above our detection limit of $\approx 10^{-4}$ ACIS-S cts s$^{-1}$  
($L_{\rm x} \approx 6 \times 10^{36}$\lum\ for the spectral 
model assumed in Sect.~2.5), 
7 out of the 74 point sources are supersoft ($kT \simlt 100$ eV), 
another 7 are in the soft subgroup. Of the remaining 60 sources, 
about half are XRBs and the other half are likely to be background AGN.
Hence, supersoft sources represent about 15\% of the true galactic 
sources brighter than our detection limit, and soft sources another 15\%.
Among the sources brighter than $3 \times 10^{-4}$ ACIS-S cts s$^{-1}$, 
none are supersoft, 2 are soft and 44 are consistent with 
XRB colours (20 of them may be AGN). Hence, the fraction of soft sources
is less than 8\%, much lower than in more actively star-forming 
late-type spiral galaxies. For example, 
we can compare M\,74 with M\,83 (Soria \& Wu 2003), 
taking into account that the distance to M\,83 is $\approx 1/2$ of 
the distance to M\,74 (de Vaucouleurs et al.~1991; Thim et al.~2003).
In M\,83, soft sources represent more than 40\% of the X-ray point sources 
brighter than what would be the detection limit in our M\,74 study, 
and 25\% of the sources in the brighter subgroup.
This is consistent with the interpretation of the soft sources 
as candidate X-ray SNRs: the number of detected SNRs 
at a given luminosity is related 
to the recent star-formation rate in a galaxy, 
as well as to other factors such as the density 
of the interstellar medium. A more comprehensive 
study of these issues is left to further work.

\section{Conclusion}

We investigated the X-ray source population in the late-type spiral M\,74.
We took a new \xmm\ observation in 2003 January to study the evolution 
of the rare Type Ic SN~2002ap a year after the event, and monitor the 
variability of two bright BH candidates. In addition, we used 
archival {\it Chandra} ACIS observations from 2001 June and October
to determine the luminosity and colour distribution 
of all the other discrete sources. 

We have more accurately determined 
the X-ray colours and luminosity of SN~2002ap five days after 
the event, and put an upper limit to its luminosity a year later.
Unlike other well-studied core-collapse SNe 
(eg, SN~1999em: Pooley et al.~2002), 
SN~2002ap was not dominated by hard X-ray emission 
in its early stages. The X-ray colours observed in 2002 February 
were soft, suggesting that the dominant component 
was optically-thin thermal emission at $kT < 1$ keV. 
Assuming an absorbing column density $\simlt 10^{21}$ cm$^{-2}$, 
we estimate an emitted luminosity of $\approx 5 \times 10^{37}$ 
erg s$^{-1}$ in the $0.3$--$12$ keV 
band, on 2002 February 2. The emitted luminosity 
in the $2$--$12$ keV band is at least an order of magnitude lower.
After 11 months, the source was no longer detected, 
implying that its luminosity had decreased 
by at least a factor of seven. We argue that 
the prompt soft X-ray emission was coming from 
the reverse shock, and that this is consistent with a low mass-loss 
rate and high stellar-wind velocity from the progenitor, 
as expected in Type Ic events. Unlike for other Type Ic 
events such as SN~1998bw, there is no evidence of 
relativistic ejecta.

Seventy-four discrete X-ray sources (not including two obvious 
foreground stars) were detected inside the $D_{25}$ 
ellipse in the combined {\it Chandra} observations, 
with a detection limit of $\approx 6 \times 10^{36}$\lum\ 
(completeness limit $\approx 2 \times 10^{37}$\lum).
After subtracting the estimated background AGN contribution, 
the luminosity distribution of the discrete sources 
is well modelled by a simple power-law of slope $\approx -0.9$, similar 
to the value inferred for other moderately active 
star-forming spiral disks. About 15\% of the true M\,74 
sources have soft colors consistent with thermal SNRs: 
they are all fainter than a {\it Chandra} ACIS-S 
count rate of $\approx 5 \times 10^{-4}$ cts s$^{-1}$, 
corresponding to emitted luminosities $\approx 2 
\times 10^{37}$ erg s$^{-1}$ (for a $kT = 0.6$ keV 
thermal plasma model and $n_{\rm H} = 10^{21}$ cm$^{-2}$). 
All brighter sources are likely XRBs.

We have studied the temporal and spectral behaviour 
of the two brightest X-ray sources previously found in M\,74. 
XMMU J013636.5$+$155036 (emitted luminosity $\approx 1.6 
\times 10^{39}$ erg s$^{-1}$ in 2002) is no longer detected in 2003 
and must now be fainter than $5 \times 10^{37}$\lum. 
The flaring source CXOU J013651.1$+$154547 has settled 
into a soft state: its power-law spectrum has photon index 
$\Gamma \approx 2.5$; any additional thermal component 
is constrained to temperatures $\simlt 0.25$ keV. 
Its luminosity shows only little variations around an average value  
$\approx 8 \times 10^{38}$\lum. In the 21 ks of our 
2003 January observation we did not find any of the hard 
flares (peak luminosity $\approx 10^{40}$\lum) 
detected in 2001--2002 (Krauss et al.~2003). 
The nature of the super-Eddington bursts remains unclear: 
we argue that they could either be due to beamed emission 
in a microquasar/jet scenario (as proposed by Krauss et al.~2003), 
or be analogous to Type-II bursts in neutron star XRBs, 
ie due to episodes of spasmodic/intermittent accretion 
in addition to a steady sub-Eddington component.

\begin{acknowledgements}

  We thank Norbert Schartel for his assistance in the \xmm\ ToO observation, 
Miriam Krauss for sending us her latest preprint, Mat Page and Kinwah Wu for 
useful discussions, and an anonymous referee for his/her suggestions.
     
\end{acknowledgements}

\appendix
\section{{\it Chandra} source list and count rates}
   \begin{table*}[h]
      \caption[]{Source ID and average count rates (total and in three narrow bands) 
of the discrete X-ray sources detected by 
{\it Chandra} ACIS-S in the combined 2001 June--October observation. 
The count rates are in units of $10^{-4}$ cts s$^{-1}$. Errors are $1\sigma$, with $3\sigma$ upper limits 
for sources not detected in a certain band.}
         \label{}
        $$ \begin{array}[width=textwidth]{lrrrr}
            \hline
            \hline
            \noalign{\smallskip}
           {\mathrm{IAU~Name}}  
		 & ~~{\mathrm{Count}}~{\mathrm{rate}}
		 & ~~{\mathrm{Count}}~{\mathrm{rate}}
		 & ~~{\mathrm{Count}}~{\mathrm{rate}}
		 & ~~{\mathrm{Count}}~{\mathrm{rate}}\\
		 & ~~(0.3-8)~{\mathrm{keV}}
		 & ~~(0.3-1)~{\mathrm{keV}}
		 & ~~~(1-2)~{\mathrm{keV}}
		 & ~~(2-8)~{\mathrm{keV}}\\
            \noalign{\smallskip}
            \hline\\
{\mathrm{CXOU~J013623.5}}+{\mathrm{154458}}~~	&  	5.7\pm 1.2 	&  0.8\pm0.4 & 	2.3\pm0.7 &	1.8\pm0.6 \\
{\mathrm{CXOU~J013623.6}}+{\mathrm{154309}}~~ 	&	8.5\pm1.5 &	1.3\pm0.4 &	4.0\pm1.0 & 	3.4\pm0.9 \\
{\mathrm{CXOU~J013625.1}}+{\mathrm{154859}}~~	&      17.9\pm1.4 &     2.9\pm0.6 &     10.1\pm1.1 &  	4.7\pm0.7 \\
{\mathrm{CXOU~J013626.6}}+{\mathrm{154458}}~~	&      1.9\pm0.5 &    	0.6\pm0.1 &    	0.7\pm0.3 &	0.5\pm0.2 \\
{\mathrm{CXOU~J013626.7}}+{\mathrm{154304}}~~	&      5.6\pm1.2 & 	0.1\pm0.1 &	2.2\pm0.7 &	3.2\pm0.9 \\
{\mathrm{CXOU~J013627.8}}+{\mathrm{154752}}~~	&      33.4\pm1.9 &     12.1\pm1.2 &     12.6\pm1.2 &   8.7\pm1.0 \\
{\mathrm{CXOU~J013628.7}}+{\mathrm{154859}}~~	&      1.7\pm0.5 &    	0.1\pm0.1 &    	0.1\pm0.1 &     1.1\pm0.4 \\
{\mathrm{CXOU~J013629.0}}+{\mathrm{154319}}~~	&      1.2\pm0.4 &	0.2\pm0.1 &    	0.5\pm0.2 &    	0.5\pm0.2 \\
{\mathrm{CXOU~J013629.5}}+{\mathrm{154656}}~~	&      4.3\pm0.7 &	<0.1 &   	<0.1 &     	4.3\pm0.7 \\
{\mathrm{CXOU~J013630.0}}+{\mathrm{154855}}~~	&      3.2\pm0.6 &     	2.5\pm0.5 &    	0.6\pm0.2 &   	<0.1 \\
{\mathrm{CXOU~J013630.1}}+{\mathrm{154520}}~~	&      8.5\pm1.0 &    	0.2\pm0.1 &     4.7\pm0.7 &     3.7\pm0.7 \\
{\mathrm{CXOU~J013630.4}}+{\mathrm{154519}}~~	&      9.4\pm1.1 &     	2.7\pm0.6 &     3.9\pm0.7 &     2.9\pm0.6 \\
{\mathrm{CXOU~J013631.1}}+{\mathrm{154458}}~~	&      4.5\pm0.8 &   	<0.1 &     	1.9\pm0.5 &     2.2\pm0.5 \\
{\mathrm{CXOU~J013631.7}}+{\mathrm{154848}}~~	&      15.2\pm1.3 &     3.4\pm0.6 &     8.3\pm1.0 &     3.4\pm0.6 \\
^{a}{\mathrm{CXOU~J013631.9}}+{\mathrm{154507}}~~&      12.2\pm1.2 &   	10.1\pm1.1 &   	1.9\pm0.5 &    	0.2\pm0.1 \\
{\mathrm{CXOU~J013635.2}}+{\mathrm{154657}}~~	&      3.1\pm0.6 &	0.2\pm0.1 &     1.1\pm0.4 &     1.8\pm0.5 \\
{\mathrm{CXOU~J013635.3}}+{\mathrm{154711}}~~	&      4.0\pm0.7 &     	1.6\pm0.4 &     1.9\pm0.5 &    	0.3\pm0.2 \\
{\mathrm{CXOU~J013635.3}}+{\mathrm{154227}}~~	&      2.2\pm0.6 &   	<0.1 &    	0.5\pm0.2 &     1.7\pm0.5 \\
{\mathrm{CXOU~J013635.4}}+{\mathrm{154953}}~~	&      1.7\pm0.5 &    	0.4\pm0.2 &    	0.3\pm0.1 &    	1.0\pm0.3 \\
{\mathrm{CXOU~J013635.7}}+{\mathrm{154556}}~~	&      7.4\pm0.9 &     	2.2\pm0.5 &     2.9\pm0.6 &     2.2\pm0.5 \\
{\mathrm{CXOU~J013637.5}}+{\mathrm{155030}}~~	&      1.3\pm0.4 &  	<0.1 &    	0.8\pm0.3 &    	0.5\pm0.2 \\
{\mathrm{CXOU~J013637.6}}+{\mathrm{154717}}~~	&      4.6\pm0.7 &     	1.9\pm0.5 &     1.9\pm0.5 &    	0.7\pm0.3 \\
{\mathrm{CXOU~J013637.7}}+{\mathrm{154740}}~~	&      9.0\pm1.0 &     	2.7\pm0.6 &     3.4\pm0.6 &     2.9\pm0.6 \\
{\mathrm{CXOU~J013637.9}}+{\mathrm{154749}}~~	&      2.1\pm0.5 &    	0.6\pm0.2 &     1.0\pm0.4 &    	0.3\pm0.2 \\
{\mathrm{CXOU~J013638.8}}+{\mathrm{154404}}~~	&      1.3\pm0.4 &     	1.1\pm0.4 &   	<0.2 & 		<0.2  \\  
{\mathrm{CXOU~J013639.0}}+{\mathrm{154755}}~~	&      5.7\pm0.8 &     	2.0\pm0.5 &     2.1\pm0.5 &     1.7\pm0.4 \\
{\mathrm{CXOU~J013639.1}}+{\mathrm{154309}}~~	&      35.6\pm2.1 &     11.0\pm1.2 &    16.9\pm1.4 &   	7.4\pm1.0 \\
{\mathrm{CXOU~J013639.2}}+{\mathrm{154600}}~~	&      3.0\pm0.6 &     	1.5\pm0.4 &    	1.0\pm0.3 &    	0.1\pm0.1 \\
{\mathrm{CXOU~J013639.3}}+{\mathrm{154744}}~~	&      10.2\pm1.1 &   	<0.1 &     	2.3\pm0.5 &     8.1\pm0.9 \\
{\mathrm{CXOU~J013639.4}}+{\mathrm{154905}}~~	&      3.3\pm0.6 &     	2.0\pm0.5 &    	0.9\pm0.3 &    	0.3\pm0.2 \\
{\mathrm{CXOU~J013639.6}}+{\mathrm{154830}}~~	&      2.4\pm0.6 &     	1.6\pm0.4 &    	0.5\pm0.2 &    	0.1\pm0.1 \\
{\mathrm{CXOU~J013639.6}}+{\mathrm{154749}}~~	&      1.2\pm0.4 &     	1.0\pm0.4 &    	0.1\pm0.1 &   	<0.1 \\
{\mathrm{CXOU~J013640.0}}+{\mathrm{154625}}~~	&      4.1\pm0.7 &     	2.5\pm0.5 &     1.4\pm0.4 &   	<0.1 \\
{\mathrm{CXOU~J013640.3}}+{\mathrm{154735}}~~	&      1.4\pm0.4 &     	1.2\pm0.4 &   	<0.1 &   	<0.1 \\
{\mathrm{CXOU~J013640.5}}+{\mathrm{154928}}~~	&      6.9\pm0.9 &    	0.9\pm0.3 &     3.1\pm0.6 &     2.8\pm0.6 \\
{\mathrm{CXOU~J013640.6}}+{\mathrm{154713}}~~	&      7.0\pm0.9 &     	2.5\pm0.5 &     2.2\pm0.5 &     2.2\pm0.5 \\
{\mathrm{CXOU~J013640.6}}+{\mathrm{154647}}~~	&      4.0\pm0.7 &     	1.0\pm0.4 &     1.7\pm0.4 &     1.2\pm0.4 \\
{\mathrm{CXOU~J013641.0}}+{\mathrm{154705}}~~	&      1.7\pm0.5 &    	0.4\pm0.2 &    	0.6\pm0.2 &    	0.4\pm0.2 \\
{\mathrm{CXOU~J013641.3}}+{\mathrm{154650}}~~	&      1.0\pm0.4 &    	0.3\pm0.1 &    	0.4\pm0.2 &    	0.2\pm0.1 \\
{\mathrm{CXOU~J013641.6}}+{\mathrm{155117}}~~	&      2.0\pm0.5 &  	<0.1 &    	0.1\pm0.1 &       1.8\pm0.5 \\
{\mathrm{CXOU~J013641.6}}+{\mathrm{154552}}~~	&      32.9\pm1.9 &     7.9\pm0.9 &     14.5\pm1.3 &  	10.5\pm1.1 \\
           \hline\\
         \end{array} $$
\begin{list}{}{}
\item[$^a$] Foreground star
\end{list}
\end{table*}

   \begin{table*}
         \label{}
        $$ \begin{array}[width=textwidth]{lrrrr}
            \hline
            \hline
            \noalign{\smallskip}
           {\mathrm{IAU~Name}}  
		 & ~~{\mathrm{Count}}~{\mathrm{rate}}
		 & ~~{\mathrm{Count}}~{\mathrm{rate}}
		 & ~~{\mathrm{Count}}~{\mathrm{rate}}
		 & ~~{\mathrm{Count}}~{\mathrm{rate}}\\
		 & ~~(0.3-8)~{\mathrm{keV}}
		 & ~~(0.3-1)~{\mathrm{keV}}
		 & ~~~(1-2)~{\mathrm{keV}}
		 & ~~(2-8)~{\mathrm{keV}}\\
            \noalign{\smallskip}
            \hline\\
^b{\mathrm{CXOU~J013641.7}}+{\mathrm{154701}}~~	&      18.7\pm1.5 &     6.7\pm0.9 &     7.3\pm0.9 &     4.8\pm0.7 \\
{\mathrm{CXOU~J013641.9}}+{\mathrm{154721}}~~	&      1.1\pm0.4 &     	1.2\pm0.4 &  	<0.1 &  	<0.1 \\
{\mathrm{CXOU~J013642.0}}+{\mathrm{154430}}~~	&      1.3\pm0.4 &     	1.2\pm0.4 &    	0.1\pm0.1 &    	0.1\pm0.1 \\
{\mathrm{CXOU~J013642.0}}+{\mathrm{154857}}~~	&      2.9\pm0.6 &    	0.6\pm0.3 &     1.3\pm0.4 &    	0.8\pm0.3 \\
{\mathrm{CXOU~J013642.4}}+{\mathrm{154701}}~~	&      2.2\pm0.5 &     	1.6\pm0.4 &    	0.3\pm0.1 &    	0.1\pm0.1 \\
{\mathrm{CXOU~J013643.2}}+{\mathrm{154709}}~~	&      5.1\pm0.8 &    	0.6\pm0.2 &     2.0\pm0.5 &     2.3\pm0.5 \\
{\mathrm{CXOU~J013643.6}}+{\mathrm{154742}}~~	&      4.5\pm0.7 &     	1.0\pm0.3 &     1.8\pm0.4 &     1.0\pm0.4 \\
{\mathrm{CXOU~J013643.8}}+{\mathrm{154357}}~~	&      5.4\pm0.8 &    	1.0\pm0.3 &     2.4\pm0.5 &     1.8\pm0.5 \\
{\mathrm{CXOU~J013643.8}}+{\mathrm{155022}}~~	&      4.1\pm0.7 &    	0.9\pm0.3 &     1.8\pm0.5 &     1.5\pm0.4 \\
{\mathrm{CXOU~J013644.0}}+{\mathrm{154908}}~~	&      2.7\pm0.6 &    	0.4\pm0.2 &     1.3\pm0.4 &    	0.9\pm0.3 \\
{\mathrm{CXOU~J013644.1}}+{\mathrm{154818}}~~	&      31.1\pm1.9 &     5.3\pm0.8 &     16.5\pm1.3 &     9.4\pm1.0 \\
{\mathrm{CXOU~J013644.3}}+{\mathrm{154629}}~~	&      4.0\pm0.7 &     	3.2\pm0.6 &    	0.5\pm0.2 &    	0.3\pm0.2 \\
{\mathrm{CXOU~J013644.6}}+{\mathrm{154848}}~~	&      1.4\pm0.5 &     	1.1\pm0.4 &   	0.1\pm0.1 &    	0.1\pm0.1 \\
{\mathrm{CXOU~J013644.9}}+{\mathrm{154546}}~~	&      5.6\pm0.9 &     	1.7\pm0.4 &     1.4\pm0.4 &     2.0\pm0.5 \\
{\mathrm{CXOU~J013645.1}}+{\mathrm{154837}}~~	&      1.4\pm0.4 &    	0.1\pm0.1 &    	0.5\pm0.2 &    	0.8\pm0.2 \\
{\mathrm{CXOU~J013645.2}}+{\mathrm{154747}}~~	&      3.7\pm0.7 &     	1.5\pm0.4 &     1.3\pm0.4 &    	0.2\pm0.2 \\
{\mathrm{CXOU~J013645.3}}+{\mathrm{154910}}~~	&      2.7\pm0.6 &    	0.5\pm0.2 &     1.1\pm0.4 &     1.0\pm0.4 \\
{\mathrm{CXOU~J013646.0}}+{\mathrm{154422}}~~	&      11.0\pm1.2 &     1.9\pm0.5 &     5.8\pm0.8 &     3.1\pm0.6 \\
{\mathrm{CXOU~J013646.1}}+{\mathrm{154842}}~~	&      3.1\pm0.6 &     	3.0\pm0.6 &    	0.1\pm0.1 &   	<0.1 \\
{\mathrm{CXOU~J013646.6}}+{\mathrm{154611}}~~	&     0.7\pm0.3 &    	0.7\pm0.3 &   	<0.1 &		<0.1 \\
^c{\mathrm{CXOU~J013647.4}}+{\mathrm{154745}}~~	&      35.7\pm2.0 &     28.5\pm1.8 &     5.9\pm0.8 &    0.8\pm0.2 \\
{\mathrm{CXOU~J013648.0}}+{\mathrm{154445}}~~	&      1.8\pm0.5 &    	0.3\pm0.2 &     1.3\pm0.4 &    	0.2\pm0.1 \\
{\mathrm{CXOU~J013648.8}}+{\mathrm{154653}}~~	&      3.9\pm0.7 &    	0.7\pm0.3 &     1.6\pm0.4 &     1.5\pm0.4 \\
{\mathrm{CXOU~J013649.1}}+{\mathrm{154527}}~~	&      5.2\pm0.8 &    	0.1\pm0.1 &    	0.9\pm0.3 &     3.8\pm0.7 \\
{\mathrm{CXOU~J013650.0}}+{\mathrm{154931}}~~	&      7.0\pm0.9 &     	1.2\pm0.4 &     2.2\pm0.5 &     2.8\pm0.6 \\
{\mathrm{CXOU~J013650.2}}+{\mathrm{154915}}~~	&      9.6\pm1.1 &     	2.7\pm0.6 &     3.9\pm0.7 &     2.1\pm0.5 \\
{\mathrm{CXOU~J013650.3}}+{\mathrm{155117}}~~	&      1.3\pm0.4 &   	<0.1 &	    	0.6\pm0.2 &    	0.6\pm0.2 \\
{\mathrm{CXOU~J013651.1}}+{\mathrm{154547}}~~	&      131.2\pm3.9 &    60.5\pm2.6 &    48.0\pm2.3 &    23.3\pm1.7 \\
{\mathrm{CXOU~J013651.2}}+{\mathrm{154339}}~~	&      13.5\pm1.4 &     2.9\pm0.6 &     8.1\pm0.9 &     1.9\pm0.5 \\
{\mathrm{CXOU~J013651.8}}+{\mathrm{155135}}~~	&      2.4\pm0.6 &    	0.6\pm0.3 &    	0.7\pm0.3 &    	0.8\pm0.4 \\
{\mathrm{CXOU~J013652.3}}+{\mathrm{154737}}~~	&      9.4\pm1.1 &     	2.6\pm0.6 &     4.3\pm0.7 &     2.4\pm0.6 \\
{\mathrm{CXOU~J013653.0}}+{\mathrm{155008}}~~	&      6.2\pm0.9 &     	1.3\pm0.4 &     3.2\pm0.6 &     1.1\pm0.4 \\
{\mathrm{CXOU~J013653.0}}+{\mathrm{154510}}~~	&      1.7\pm0.5 &    	0.1\pm0.1 &     0.6\pm0.3 &    	0.8\pm0.3 \\
{\mathrm{CXOU~J013653.8}}+{\mathrm{154537}}~~	&      1.4\pm0.5 &   	<0.1 &	         0.1\pm0.1 &	1.3\pm0.4 \\
{\mathrm{CXOU~J013654.4}}+{\mathrm{154539}}~~	&     1.4\pm0.5 &    	0.4\pm0.2 &    	0.5\pm0.3 &    	0.3\pm0.2 \\
{\mathrm{CXOU~J013659.3}}+{\mathrm{154631}}~~	&      5.6\pm0.9 &    	0.4\pm0.2 &     1.4\pm0.4 &     4.2\pm0.7 \\
           \hline\\
         \end{array} $$
\begin{list}{}{}
\item[$^b$] Nucleus of M\,74
\item[$^c$] Foreground star
\end{list}
\end{table*}

\end{document}